# Angular dependence of the interlayer coupling at the interface between two dimensional materials 1T-PtSe$_2$ and graphene

P. Mallet[1], F. Ibrahim[2], K. Abdukayumov[2], A. Marty[2], C. Vergnaud[2], F. Bonell[2], M. Chshiev[2], M. Jamet[2] and J-Y Veuillen[1,*]

[1] Univ. Grenoble Alpes, CNRS, Grenoble INP, Institut NEEL, 38000 Grenoble, France

[2] Univ. Grenoble Alpes, CEA, CNRS, Grenoble INP, IRIG-SPINTEC, 38000 Grenoble, France

* : jean-yves.veuillen@neel.cnrs.fr

**Abstract:**

We present a study by Scanning Tunneling Microscopy, supported by ab initio calculations, of the interaction between graphene and monolayer (semiconducting) PtSe$_2$ as a function of the twist angle θ between the two layers. We analyze the PtSe$_2$ contribution to the hybrid interface states that develop within the bandgap of the semiconductor to probe the interaction. The experimental data indicate that the interlayer coupling increases markedly with the value of θ, which is confirmed by ab initio calculations. The moiré patterns observed within the gap are consistent with a momentum conservation rule between hybridized states, and the strength of the hybridization can be qualitatively described by a perturbative model.

Proximity effect allows transferring some electronic properties at the interface between two materials without dramatically affecting their individual structure. This phenomenon is thought to be especially relevant for few layers thick materials where even short range interactions can impact the properties of the whole sample [1]. Therefore, proximity effect represents an efficient strategy for tuning the properties of van der Waals heterostructures [2]. For instance, it has been exploited to imprint a spin dependent electronic structure in graphene when brought in contact with semi-conducting (SC) or insulating two dimensional (2D) layers exhibiting either strong spin-orbit coupling or magnetic ordering [1-4]. Inducing either a significant spin-orbit interaction [5-16] and/or a large magnetic exchange [17-25] in graphene is highly desired for developing spintronics based on 2D materials [1, 3, 4, 26, 27]. Considering the relevance of the proximity effect, it is crucial to determine which factors drive the strength of the coupling between a SC 2D layer and graphene. One important parameter is the rotation angle θ between the atomic lattices of the two layers. Recent theoretical studies [28-32] have indeed revealed that the amplitude and even the sign of the proximity effect can strongly depend on θ. Exploiting this angular dependence is the key aspect of the rapidly developing field of "twistronics" [33, 34].

In this work, we have investigated the basic mechanisms underlying the interaction between graphene and a SC 2D transition metal dichalcogenide (TMD) layer by means of scanning tunneling microscopy (STM) and spectroscopy (STS), complemented by ab initio simulations. Our approach consists in analyzing the electronic states that appear within the SC bandgap on the TMD side of the heterostructure as a consequence of the interfacial coupling. We thus probe the component in the PtSe$_2$ layer of the states formed within the TMD bandgap by hybridization with the low energy states of graphene. This includes the states close to the Dirac point (DP) of (bilayer) graphene, that is located around 0.3 eV below the Fermi level in the samples we use [35]. This hybridization is precisely the mechanism that generates the spin dependent proximity effect [1, 3, 4, 6, 17, 18, 28, 29, 31, 32], and our approach allows to probe it directly. We have analyzed domains corresponding to different values of θ. Our data show evidence for the angular dependence of the interfacial coupling and support the perturbative model presented in Refs. 28, 29 and 36. Notice that our approach is different from the one reported recently [37], where the proximity effect was studied by probing quantum interference patterns from the graphene side.

As a SC TMD we have chosen monolayer (1L) $PtSe_2$, which has a bandgap of about 2 eV. $1T-PtSe_2$ actually shows a comparatively large interlayer interaction in the TMD family [38, 39, 40]. If this property is preserved at the interface with graphene, it would lead to a large hybridization and thus to a detectable signal within the TMD bandgap in the STM measurements. Although we do not address this point, 1L $PtSe_2$ is also potentially interesting to induce spin dependent proximity effects in graphene, owing to a strong spin-orbit coupling [41, 42, 43] and to a long range defect-induced magnetic ordering [44].

The $PtSe_2$ layer was grown by molecular beam epitaxy (MBE) on a graphene bilayer substrate (BLG) grown on SiC [45]. The STM measurements were performed in a setup operating at 8.5K, and analyzed using the WSxM software [46]. Ab initio calculations were performed using the VASP package [47, 48; 49]. Technical details can be found in the Supplemental Material [50] (see also references [51-64] therein).

Fig. 1-a and Fig. S1 [50] show that the sample surface is mostly covered by the monolayer $PtSe_2$ (1L $PtSe_2$) phase, with domains size of the order of 10 nm. Some patches of BLG substrate remain visible, and a few bilayer $PtSe_2$ islands (not discussed here) are also present. In many domains, the $PtSe_2$ lattice is almost aligned with the BLG (see section SI2 in Ref. [50]), as already reported in samples grown on Highly Ordered Pyrolytic Graphite (HOPG) [65, 66, 67]. This leads to a quasi-(2x2) superstructure in STM images, since twice the lattice constant of $PtSe_2$ (0.375 nm) is almost equal to three times the lattice constant of graphene (0.246 nm). In such domains, the STS data (Fig. S2-d [50]) reveal a bandgap of about 1.90 eV, in agreement with previous works [65, 66, 67]. The valence band maximum (VBM) and conduction band minimum (CBM) are located close to -1.65 eV below and +0.25 eV above the Fermi level respectively. All the 1L $PtSe_2$ domains probed in this work are featureless and have the same apparent height when imaged at +1.2 V sample bias, this is about 1 eV above the CBM, as shown in Fig. 1-a and Fig. S1 [50]. However, when the sample bias is set inside the 1L $PtSe_2$ gap as in Fig. 1-b, or close to the VBM as in Fig. S3-b or in Fig. S5-a [50], their appearance changes remarkably. Triangular superstructures with periods P in the nm range appear inside the islands, whose apparent height depends on the value of P. "Large" (≥1 nm) values of P correspond to apparently higher islands. Height differences of the order of 0.1 nm are measured between domains with different periods as shown in Fig. 1-c and Fig. S3-c [50]. In the data taken at a sample bias of +1.2 V, tunneling occurs mainly towards empty states of 1L $PtSe_2$, and the images should represent the topography of this layer. Conversely, for sample bias within the TMD gap, the images should reflect the electronic structure of the 1L $PtSe_2$-BLG interface. Since free-standing 1L $PtSe_2$ has no states in this energy range, the electron states that contribute to the tunneling current within the TMD gap must originate mostly from those of the underlying BLG substrate, perturbed by the contact with 1L $PtSe_2$. One can illustrate how the BLG states are modified at the interface by considering the current vs. voltage ($I_t(V_s)$) spectra shown in Figure 1-d. The pink curve shows the spectrum measured on the bare BLG. When the tip is moved onto the 1L TMD domains, hence retracted by roughly 0.8 nm (ie by the apparent height of the layer at the bias set point of +1.2V), a significant current is still measured within the TMD bandgap (light and dark blue curves respectively), 4 to 5 order of magnitude higher than expected for an attenuation 'as in vacuum' of the tunneling current into the BLG states (Section SI3 of Ref. [50]). This implies that the graphene derived states gain an extra spectral weight within the TMD layer in the energy range of the bandgap, thus that they acquire a partial $PtSe_2$ character. Otherwise stated, there is a finite, albeit small, admixture of states originating from $PtSe_2$ into the mainly graphene-based in-gap states (Section SI3

in Ref. 50). This is a strong indication for the existence of hybridized Graphene-PtSe$_2$ electron states within the TMD bandgap. The experimental variations of the $I_t(V_s)$ curves between the two islands in Fig. 1-d and Fig. S3-d, or equivalently the variations of the apparent height of the islands in images taken with sample biases within the gap (Fig. 1-b and S3-b), indicate that the admixture of PtSe$_2$ derived states depends on the value of P. The larger current or apparent height for long period (P≥1 nm) superstructures suggests that the hybridization should be stronger in this case.

Ab initio band structure calculations confirm the existence of such hybridized states at the BLG-1L PtSe$_2$ bilayer, as shown in Fig. 1-e and 1-f (see also Fig. S11 [50]). The atomic lattices of BLG and PtSe$_2$ are aligned for this calculation. Therefore, due to the (3x3) BLG supercell used, both +K and –K points of the Brillouin zone (BZ) of BLG are backfolded at Γ. The calculated band structure of the heterostructure is projected on the BLG and on the PtSe$_2$ layers in Fig. 1-e and Fig. 1-f respectively. Within the TMD bandgap at Γ (and even up to +0.75 eV) one clearly recognizes the characteristic band structure of bare BLG in Fig. 1-e (heavy red line). The same features are found, although with a much smaller weight, in Fig. 1-f (light blue lines inside the pink rectangle). This calculation shows that, inside the TMD bandgap, the eigenstates, which mainly originate from the bare BLG states, acquire a small weight on the orbitals of the Pt and Se atoms. Therefore, hybrid states with a small admixture of PtSe$_2$ form within the bangap at the TMD/BLG interface. Similar results were found for the interface between 1L PtSe$_2$ and monolayer graphene, see Fig. S11.

In Ref. 50 (section SI4) we show that the period (P) of the triangular superstructures observed inside the 1L-PtSe$_2$ islands is related to the rotation angle θ between the crystallographic directions of the 1L PtSe$_2$ and of the BLG substrate. The value of θ can be evaluated directly from images with atomic resolution on the bare BLG and on the TMD islands (Fig. S5 [50]). The superstructures are thus Moiré Patterns (MPs) resulting from this disorientation. Figure 2-a, 2-c and 2-e show atomic resolution images of domains with different values of θ. Their Fourier transfoms (FTs) are displayed in Figure 2-b, 2-d and 2-f. For the aligned domain with θ= 0° (Fig. 2-a and 2-b) one identifies readily the (2x2) superstructure. The first order spots of the superstructure are labelled as "MP1" (pink triangles) in Fig. 2-b. Strong second order spots, indicated by the white arrows, are also present. For a small but finite rotation angle θ≈4°, the real space image (Fig. 2-c) becomes more complex. The corresponding FT (Fig. 2-d) reveal two sets of spots, marked as MP1 (pink triangles) and MP2 (orange triangles), both corresponding to the superstructure. We ascribe the spots labelled MP1 to a Moiré pattern based on the difference between the (shortest) reciprocal vectors of PtSe$_2$ and of the graphene lattices [68, 69], as shown in Fig. 2-g. It coincides with the (2x2) superstructure for θ=0°. The second series of spots, labelled MP2, also corresponds to a MP constructed on the difference between the reciprocal vector of graphene (**a\***$_{Gr}$) and the sum of the reciprocal vectors of PtSe$_2$ (**a\***$_{TMD}$+**b\***$_{TMD}$) lattices, see Fig. 2-g (plus symmetry related vectors). It coincides with the second order spots of the (2x2) superstructure for θ=0° (white arrows in Fig. 2-b). For large rotation angles (θ≈20°, Fig. 2-e and 2-f) we identify one set of spots which correspond to the MP2 Moiré pattern. The computed variations of the structural parameters of MP1 and MP2 as functions of θ are displayed in Fig. S9. We find a good agreement between the period and the orientation of the superstructures obtained from the analysis of the FTs of the images and the values computed from the direct measurement of θ. Additional data on that point are presented in Ref. 50. These observations strongly support our interpretation of the superstructure in terms of those two Moiré patterns (MP1 and MP2). A very important point for the purpose of the present work is that MP2 is detected for all values of θ (assuming that the second order spot of the (2x2) for θ=0° correspond to MP2). Figures 2, S7 and S8 suggest that MP1 is strongly attenuated for θ≥9°, especially at large negative bias, whereas MP2 is present at any angle and for all sample biases. The occurence of the Moiré MP1 in our STM images is considered in section SI5 of Ref. 50, but it is not relevant for the following discussion.

Our experimental data therefore suggest the following picture for the electronic structure of the interface. Graphene (BLG) π states located within the bandgap of the TMD interact with the TMD potential, which gives rise to hybrid states having a partial $PtSe_2$ character. This "in-gap" $PtSe_2$ contribution, as probed by the STM tip, (i) shows ubiquitous spatial modulations related to the so-called MP2 moiré pattern and (ii) depends on the rotation angle between the TMD and BLG substrate, being larger for large rotation angles. Actually, these observations are consistent with the scenario proposed for the interlayer interaction at an (arbitrary) incommensurate interface [36]. This mechanism has been adapted to the case of a semiconducting TMD-graphene interface in the context of the proximity effect [28, 29]. Even though single layer graphene is considered in Ref. [28, 29], the same discussion should hold in the BLG case. The low energy states of graphene, those which are present within the 1L $PtSe_2$ bandgap, are located close to the $K_G$ (and -$K_G$) points of the graphene Brillouin zone (BZ). From Ref. [28, 29], each graphene state is coupled essentially with TMD states located at three points in the first Brillouin zone of 1L $PtSe_2$, following the momentum conservation rule [36]. This mechanism is illustrated in Figure 3-a to 3-c for different rotation angles in the case of graphene states located close to the $K_G$ point [28, 29] (the same arguments hold for the –$K_G$ point). Therefore, the hybrid interface states within the gap consist of a sum of a (dominant) graphene contribution and of a set of 1L $PtSe_2$ states with wave vectors $\mathbf{k_1}$, $\mathbf{k_2}$ and $\mathbf{k_3}$ in the 1L $PtSe_2$ first BZ (see Fig. 3). As shown in figure S10 [50], those three wave-vectors are connected by reciprocal vectors of the MP2 Moiré pattern. Consequently, for any value of θ, the squared wavefunction of each hybrid state, evaluated in the 1L $PtSe_2$ layer (i. e. disregarding the graphene component) will contain terms oscillating with the periodicity of MP2. The result is consistent with our observation of the MP2 superstructure for all orientations of the TMD layer.

Moreover, considering that a graphene state located within the bandgap can be rather far (a fraction of an eV) from the band edges of the TMD, the related hybrid state can be obtained from perturbation theory [28, 29]. The amplitude of the 1L $PtSe_2$ contributions to the hybrid state will be proportional to a matrix element and inversely proportional to the distance in energy between the graphene and the 1L $PtSe_2$ states with wavevectors $\mathbf{k_1}$, $\mathbf{k_2}$ or $\mathbf{k_3}$ [29]. This provides an explanation for the angular variation of the weight of the 1L $PtSe_2$ contribution within the gap which is observed experimentally. Indeed, as shown in Figure 3, graphene states at the $K_G$ point will be coupled to different states $\mathbf{k_i}$ (i=1 to 3) in the $PtSe_2$ layer when the rotation angle changes. One reason of the angular variation of the $PtSe_2$ weight could be that the TMD states have different Pt and Se orbital character for different values of θ [41, 66], which will modify the value of the matrix element. This effect is not intuitive and it has to be computed numerically [28, 29]. The other reason, which is the role of the distance in energy between graphene and TMD states, can be more readily evaluated considering the calculated band structure of freestanding 1L $PtSe_2$ (see Fig. 3-d), as shown in Ref. 29. The values of the vectors $\mathbf{k_i}$ of the TMD states hybridized with the graphene states at the $K_G$ point (DP) are indicated for specific values of θ in Fig. 3-d (cf. Ref. 50 for technical remarks on this figure). From the calculation of the heterojunction (TMD on BLG, see Fig 1-e and 1-f and S11), the DP of BLG is 0.4 eV below the CBM, this is close to $E_F$ in Fig. 3-d. One can thus see from Fig. 3-d that the distance in energy between a graphene state located at the $K_G$ point (DP) and the $\mathbf{k_1}$, $\mathbf{k_2}$ and $\mathbf{k_3}$ 1L $PtSe_2$ states changes markedly with θ (roughly by a factor of 2). This modifies the amplitude of the 1L $PtSe_2$ contributions to the hybrid state in the perturbative approach. Considering only this effect, one expects a weak amplitude for small values of θ (around 0°) and a stronger one for larger values of θ (around 25-30°). In the experiments, the DP of graphene is located 0.30 eV below the Fermi level [35, 50], this is 0.55 eV below the CBM [50]. The coupling of the graphene states close to the DP to the ones of the 1L $PtSe_2$ layer should thus follow the same trends as in the calculations. Therefore, we can infer that the in-gap tunneling current, or the apparent height of the islands for in-gap imaging biases, should be smaller for small angles (θ≈0°, P< 1 nm) than for larger ones (θ≥20°, P≥1 nm). This is in agreement with the experimental observations reported in Figures 1 and S3 [50], and supports the interpretation of these data in terms of hybridized states at the interface. However, considerations based only on the energy separation of the bands may be insufficient to fully describe the complexity of the systems, as revealed by ab initio studies [30, 31, 32]. It was suggested that band anticrossings may be a better way to evaluate the coupling between graphene and TMD states [32]. We adress this point in Ref. 50 (section SI6, Fig. S12) for the specific case θ=19.1°. Our ab-initio calculations also support a larger hybridization for this rotation angle than for θ=0°.

In summary, our experimental study of the 1L PtSe$_2$-BLG junction, supported by ab initio calculations, has revealed the saliant features of the interlayer interaction between the states in this twisted van der Waals heterostructure. This was achieved by analyzing the electronic states appearing inside the TMD bandgap as a result of hybridization at the interface . The momentum conservation rule gives rise to a specific Moiré pattern observed in the whole range of rotation angles. The angular dependence of the coupling strength between the layers gives rise to variations in the tunneling current measured in the TMD bandgap. Beyond the case of the 1L PtSe$_2$-BLG system, our work supports theoretical [28-32] models which propose to use the twist angle as a knob to tune the amplitude of the proximity effect in SC TMD-graphene junctions.

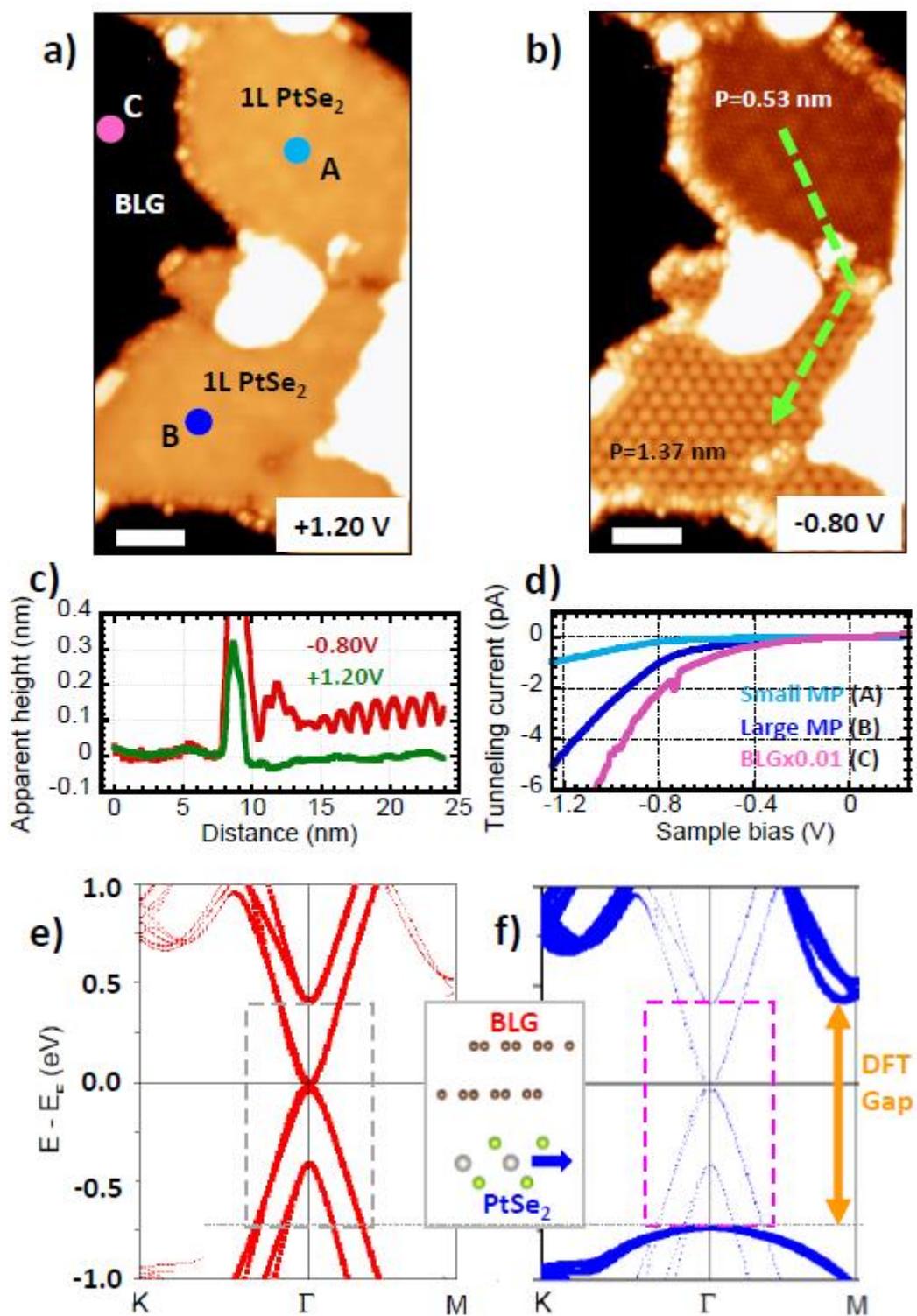

**Figure 1**

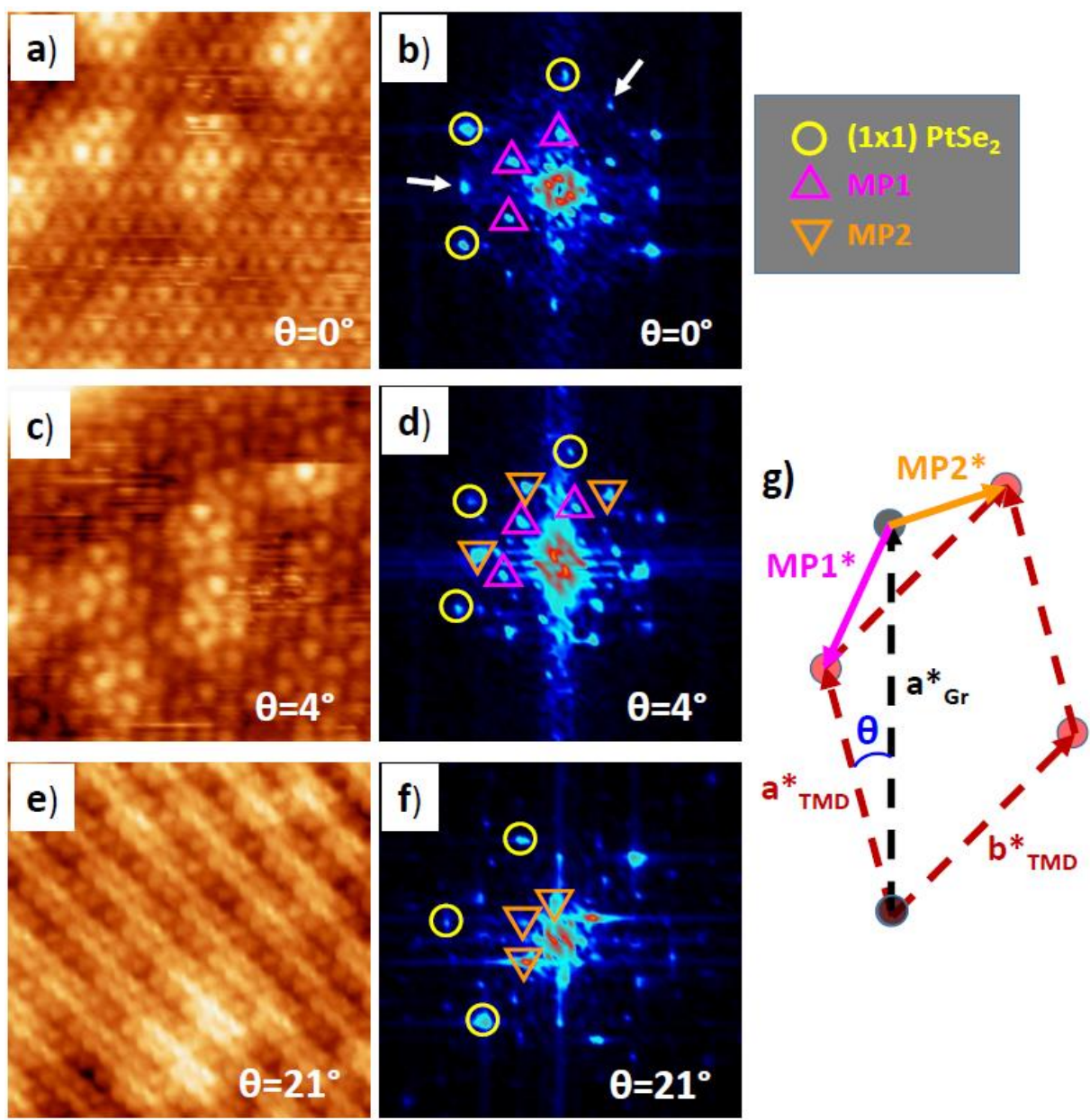

**Figure 2**

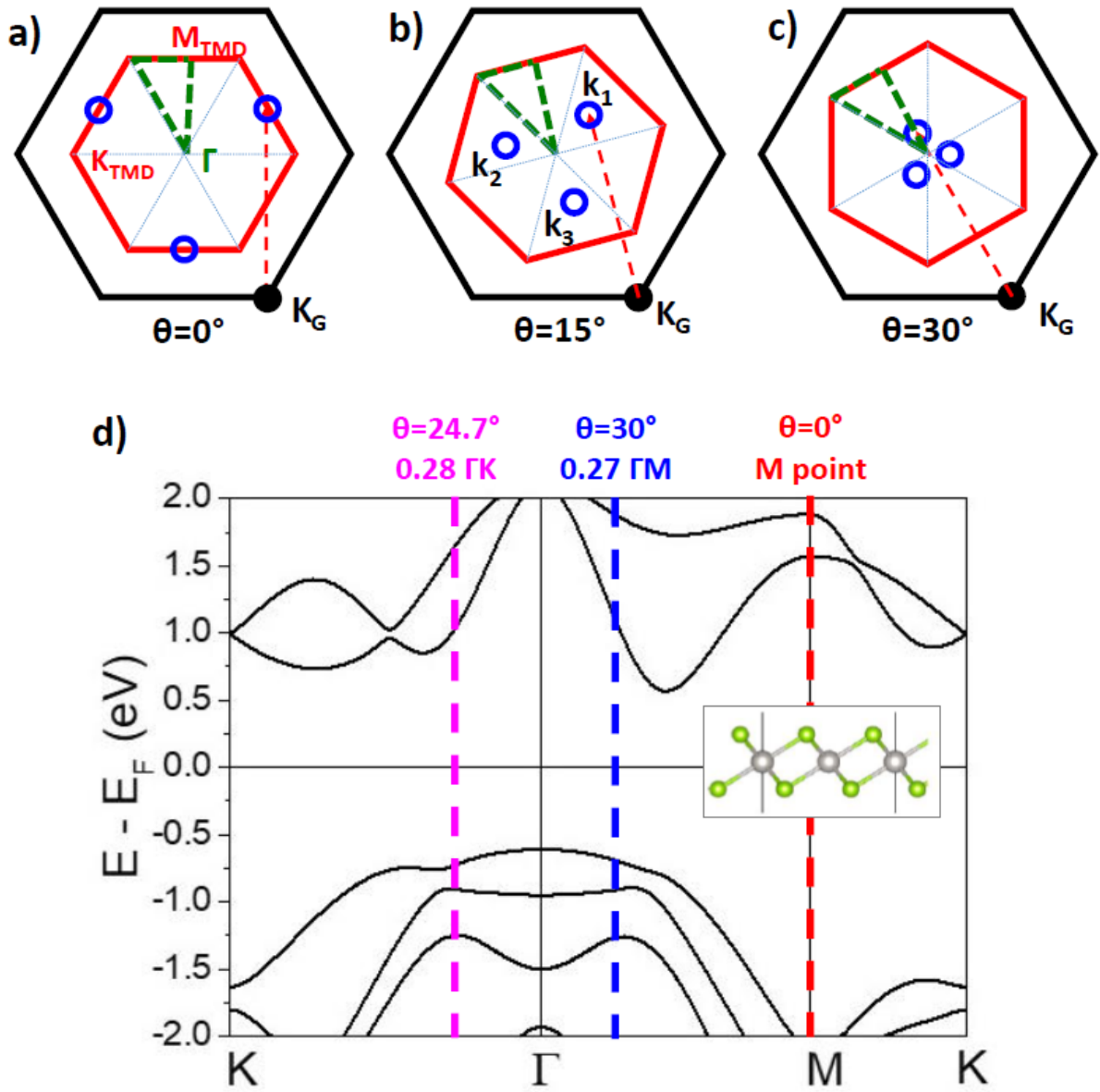

Figure 3

**Figure captions**:

Figure 1: Superstructures at the interface between 1L PtSe$_2$ and BLG. a) Constant current STM image taken with sample bias V$_s$=+1.20 V and tunneling current I$_t$=20pA. The scale bar is 5 nm. b) Image of the same area with V$_s$=-0.80 V and I$_t$=5 pA. The periods of the superstructures that appear in the upper (lower) island are 0.53 nm (1.37 nm). c) Profile of the apparent height taken on images in a) and b) along the path represented by a broken line in b). d) Tunneling spectra (I$_t$(V$_s$) curves) taken at the point labelled A (light blue curve), B (dark blue curve) and C (pink curve, on the BLG) in a). The set-point for all spectra was V$_s$=+1.20V and I$_t$=500 pA. The value of the current on the BLG is divided by 100 in the plot. e) and f) Calculated electronic structure for the interface between 1L PtSe$_2$ and BLG. The relaxed structure of the system is displayed in the central panel. The atomic lattices of BLG and 1L PtSe$_2$ are aligned (i. e. θ=0°), and the common (commensurate) unit cell is (2x2) PtSe$_2$ or (3x3) BLG. The band structure of the heterojunction is projected on the BLG in red (e) and on the 1L PtSe2 layer in blue (f) respectively. The size of the symbols (dots) in the plots is proportional to the weight of the site-projected state in the corresponding layer. The orange arrow in f) indicates the calculated TMD bandgap. The pink rectangle in f) indicates the location where hybridized states appear within the TMD bandgap (a grey rectangle is shown at the same position in e)).

Figure 2: Superstructures as Moiré patterns. a), c) and e): Constant current images with size 6x6 nm² taken on domains with different values of the twist angle θ. V$_s$= -1.80V for a) and c), Vs=+0.10 V for e). Images for other biases are shown in Fig. S8 [50]. b), d) and f): FTs of the images in a), c) and e) respectively. The first order spots of the reciprocal lattice of 1L PtSe$_2$ are indicated by yellow circles. The spots corresponding to the Moirés patterns MP1 and MP2 are indicated by pink and orange triangles respectively. g) Construction giving the reciprocal lattice vectors of the Moiré patterns MP1 and MP2 [68, 69]. a*$_{TMD}$ and a*$_{Gr}$ are the reciprocal vectors of the 1L PtSe$_2$ and BLG lattices respectively.

Figure 3: Hybridization of the graphene and TMD states at the interface. a), b) and c): Locations in the first Brillouin zone (BZ) of 1L PtSe$_2$ of the wavevectors of the TMD states coupled to the low energy BLG states for different rotation angles θ (θ=0° in a), θ=15° in b) and θ=30° in c)). The BZ of BLG (1L PtSe$_2$) are drawn in black (red) lines respectively. The construction [28, 29] that gives the wavevector k$_1$ of the TMD layer coupled to the BLG states close to K$_G$ is illustrated in a) to c), where the dashed red arrow represents a reciprocal lattice vector of 1L PtSe$_2$. k$_2$ and k$_3$ are obtained from the other equivalent K$_G$ points in the BLG BZ. d): Location of the k$_i$ (i=1 to 3) for states relative to the graphene Dirac point in the calculated band structure of freestanding 1L PtSe$_2$. The dashed vertical lines indicate the values of the k$_i$'s for specific rotation angles, where they cross the main directions (ΓM, ΓK or KM) of the BZ of 1L PtSe$_2$.

**Acknowledgements:**

The authors acknowledge the support from EU Horizon 2020 research and innovation Programme under grant agreement No 881603 (Graphene Flagship) and from the Agence Nationale de la Recherche (ANR) through the ANR-18-CE24-0007 MAGICVALLEY project.

Supplemental Material for MS entitled:

**Angular dependence of the interlayer coupling at the interface between two dimensional materials 1T-PtSe$_2$ and graphene**

by :

P. Mallet[1], F. Ibrahim[2], K. Abdukayumov[2], A. Marty[2], C. Vergnaud[2], F. Bonell[2], M. Chshiev[2], M. Jamet[2] and J-Y Veuillen[1,*]

[1] Univ. Grenoble Alpes, CNRS, Grenoble INP, Institut NEEL, 38000 Grenoble, France

[2] Univ. Grenoble Alpes, CEA, CNRS, Grenoble INP, IRIG-SPINTEC, 38000 Grenoble, France

**\*** **:** jean-yves.veuillen@neel.cnrs.fr

Content:

**SI1 : Details on the sample preparation, STM measurements and ab-initio calculations.**

**SI2 : Morphology of the layers and band onsets at the interface between 1L-PtSe$_2$ and bilayer graphene (BLG).**

**SI3 : Moiré patterns and in-gap states at the 1L-PtSe$_2$/BLG interface (additional data for Figure 1).**

**SI4 : Origin of the triangular superstructures (Moiré patterns) at the 1L-PtSe$_2$/BLG interface .**

**SI5 : Interpretation of the Moiré patterns at the 1L-PtSe$_2$/BLG interface (additional data for Figure 2).**

**SI6 : Moiré pattern and interlayer coupling (additional data for Figure 3).**

**SI1 : Details on the sample preparation, STM measurements and ab-initio calculations.**

Molecular beam epitaxy (MBE) has been used to prepare the sample studied in this work [1]. We used epitaxial bilayer graphene (BLG) on SiC(0001) as a substrate for the growth. This sample consists in a Bernal stacked (or AB stacked) bilayer of graphene on top of the so-called "buffer layer" which is a C plane strongly bound to the SiC substrate [2]. The BLG keeps the same orientation over the whole substrate surface [3] (no rotational disorder). Measurements have revealed that the BLG on SiC(0001) was heavily electron doped (in the $10^{13}$ cm$^{-2}$ range), with a Dirac point located around 0.30 eV below the Fermi level [4, 5].

The epitaxial BLG on SIC substrate was first outgassed under UHV at 800°C during 30 minutes and maintained at 280°C during the growth [1]. The base pressure in the MBE reactor was $5.10^{-10}$ mbar. Platinum was evaporated from an e-gun evaporator and selenium from an effusion cell. Quartz balance monitor was used for adjusting the rate of platinum deposition. The targeted coverage was 1 ML PtSe$_2$. The selenium pressure was measured with a hot cathode gauge, and fixed at $1.10^{-6}$ mbar. After the growth, the sample was annealed at 500°C for 20 minutes and cooled down under selenium flux. The surface crystalline order and morphology were monitored in situ by reflection high-energy electron diffraction (RHEED).

The as-grown layer was capped at room temperature with amorphous selenium in the MBE chamber to protect the sample from oxidation during transfer in air to the STM setup. The coating layer was then removed in-vacuum by a soft annealing (around 200°C), and the sample was eventually cooled down for the STM/STS measurements. All the STM/STS experiments were performed at 8.5 K with PtIr tips, using a set-up developed in the group. Local tunneling spectroscopy was achieved at fixed tip-surface separation, while maintaining the feedback-loop open. A 4 mV rms modulation with frequency 473 Hz was added to the sample bias, and the conductance was detected with the lock-in amplifier technique. The STM/STS data were processed using the WSXM software [6].

Our first-principles calculations are based on the projector-augmented wave (PAW) method [7] as implemented in the VASP package [8, 9, 10] using the generalized gradient approximation [11] and including spin-orbit coupling. The PtSe$_2$/SLG(BLG) heterostructures were constructed considering two values of the twisting angle: (1) 0° by matching 2×2 supercell of 1T-PtSe$_2$ with 3×3 supercell of graphene on top leading to a lattice mismatch ~ 1.5% and (2) 19° by matching √7×a of PtSe$_2$ with 4×4 graphene which minimizes the mismatch to 0.8%, see Fig. S12-b in Section SI6. A sufficient vacuum layer of 20 Å thickness was added to the heterostructures. The atomic coordinates were relaxed until the forces became smaller than 1 meV/Å. A kinetic energy cutoff of 550 eV has been used for the plane-wave basis set and a Γ-centered 15×15×1 k-mesh was used to sample the first Brillouin zone. To describe correctly the interaction across the interface, van der Waals forces were used with Grimme type dispersion-corrected density functional theory-D2 [12]. The site-projected band structures are calculated by summing the projections of the wavefunctions on the atomic orbitals of a specific atom as implemented in VASP [13,14].

It is important to note that performing beyond standard DFT with generalized gradient approximation functionals like PBE ones, we underestimate the bandgap of monolayer PtSe$_2$ by about 0.5 eV compared to the experimental value. However, Guan et al. [15] compared the band structure of the 1 ML PtSe$_2$/Gr heterostructure calculated using PBE and HSE06 hybrid functionals. The later shifted the PtSe$_2$ valence bands by about -0.5 eV whereas the energy position of the conduction bands is less affected. Thus, using the hybrid functional mainly results in a "rigid-band"-like shift with a minor effect on the PtSe$_2$ bands character. Moreover, although the band gap is underestimated, the dispersions of the bands (either occupied or empty) are satisfactorily described in PBE calculations. This can be appreciated for instance by comparing the experimental (valence) band structure of 1L PtSe$_2$ measured in ARPES with the computed one (see for instance Y. Wang et al. [16]). Thus, we believe that using PBE functional results in reliable band dispersions.

**SI2 : Morphology of the layers and band onsets at the interface between 1L-PtSe$_2$ and bilayer graphene (BLG).**

In this section, we show:

- In Figure S1: Large-scale images of the sample to visualize the structure of the film and the size of the islands for the monolayer (1L) phase.
- In Figure S2: Scanning Tunneling Spectroscopy data for domains of 1L-PtSe$_2$ essentially aligned with the BLG substrate, to make connection with results obtained in previous works [17, 18, 19].

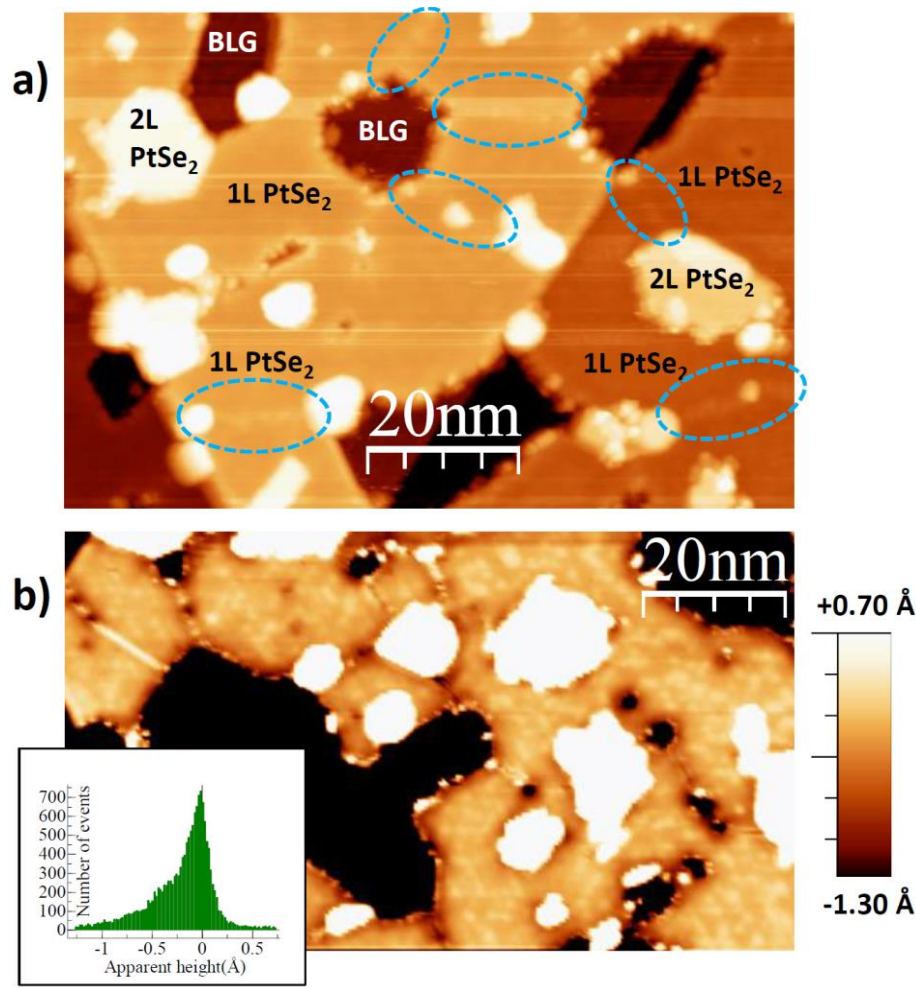

Figure S1: Sample morphology. a) Image of several substrate terraces covered with the PtSe$_2$ deposit. Size: 95x65 nm², sample bias/current: -2.50V/250pA. Monolayer PtSe$_2$ (labelled 1L PtSe$_2$) covers most of the terraces, while patches of the substrate remain visible (labelled BLG). Some domain boundaries of 1L PtSe$_2$ are indicated by blue (dashed) ovals. A few bilayer PtSe$_2$ islands (labelled 2L PtSe$_2$, not discussed there) are also present. b) Image of another area of the sample (consisting in a single BLG substrate terrace) with an enhanced contrast on 1L PtSe$_2$, which reveals the domain boundaries. Image size: 102x59 nm², sample bias/current: +1.20V/20pA. The size of the domains is in the 10 nm's range, and all of them have the same contrast. The inset shows the height histogram on the 1L PtSe$_2$ phase, which confirms that all the domains have an identical apparent height for the sample bias +1.20V.

The important point there is that all the 1L PtSe$_2$ islands have the same apparent height relative to the BLG substrate when imaged at sample bias +1.20V, this is well above the conduction band minimum (CBM), as shown below.

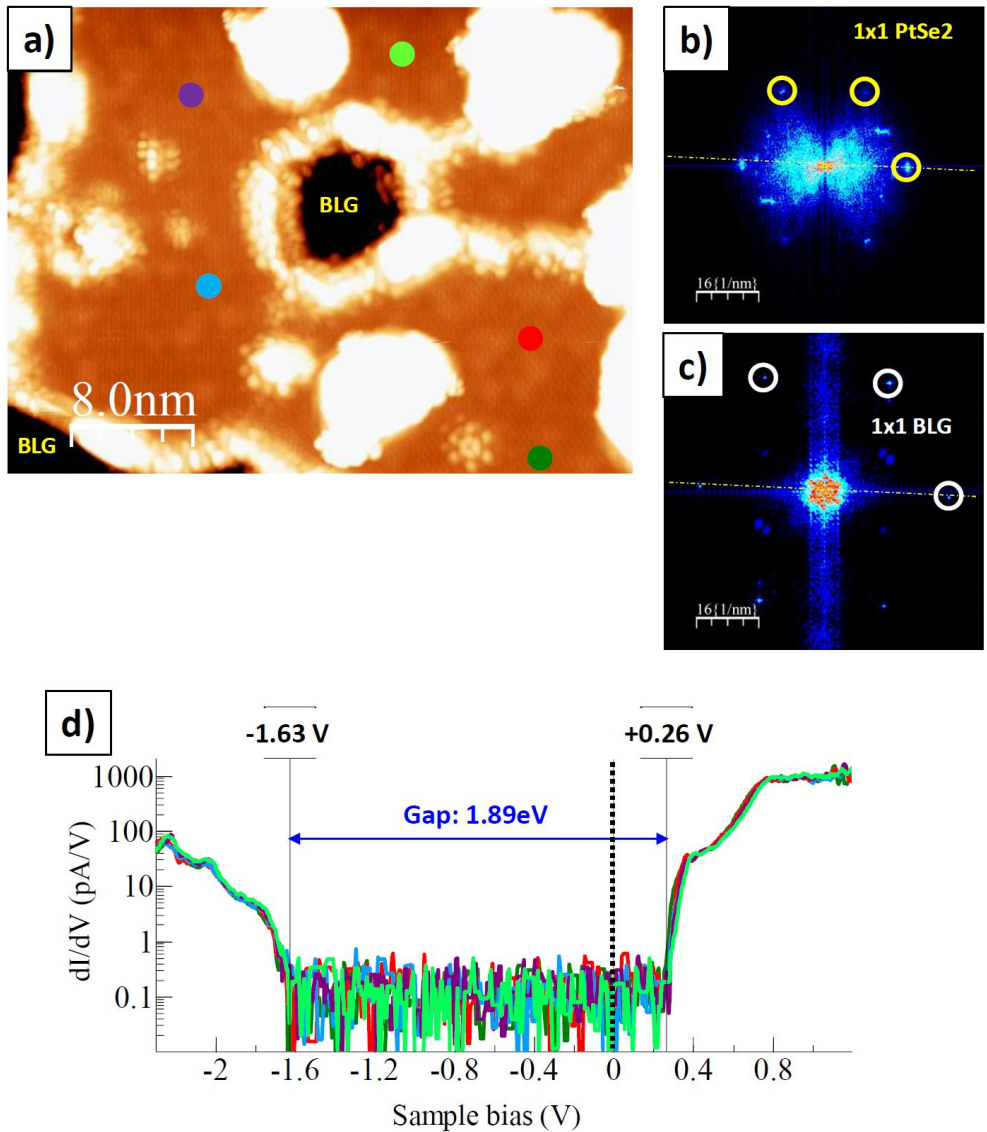

Figure S2: Electronic structure of 1L PtSe$_2$ islands aligned with the BLG substrate. a) 40x30nm² image taken on several 1L PtSe$_2$ islands on BLG, with a contrast enhanced on the PtSe$_2$ area. Sample bias/current: -0.80V/5pA. b) Fourier transform of the image in a), showing the spots of the 1x1 PtSe$_2$ lattice (yellow circles). c) Fourier transform of an image taken on the nearby BLG substrate. The spots of the 1x1 BLG lattice are indicated. It is clear that the lattices of the PtSe$_2$ islands and of the BLG substrate are almost aligned in this area (as shown by the dashed yellow line in b) and c)). d) Spectra taken at the locations of the spots with the same color in a), corresponding to different domains. Set-point bias/current: +1.20V/500pA. All spectra are quite similar. One finds the conduction band minimum (CBM) at +0.26 eV above the Fermi level (dotted black vertical line, corresponding to zero bias) and the valence band maximum (VBM) at -1.63 eV below the Fermi level. The value of the gap for 1L PtSe$_2$ on BLG is thus close to 1.90 eV. From a series of measurements we estimate that the uncertainty in the determination of the VBM (CBM) is ±0.05 eV (±0.03 eV).

Figure S2 shows the results we got for 1L PtSe$_2$ islands aligned with the BLG substrate. Our estimate for the gap width (1.90±0.05 eV from a series of measurements on different domains) is within the range of values (1.8 to 2.1 eV) reported for 1L-PtSe$_2$ grown on graphite (HOPG) [17, 18, 19]. The CBM is closer to the Fermi level here than for HOPG substrates. This is expected from the difference in the work function between BLG and graphite [20], assuming a weak Fermi level pinning [21, 22].

**SI3 : Moiré patterns and in-gap states at the 1L-PtSe$_2$/BLG interface (additional data for Figure 1).**

In this section, we present:

- In Figure S3 a set of data essentially similar to the one of Figure 1, but taken at another location on the sample.
- In Figure S4 additional measurements, which allow for estimating the order of magnitude of direct tunneling into the bare BLG states from above the 1L PtSe$_2$ island of Fig. 1.
- In Figure S5 an evidence for the existence of hybridized graphene-TMD states within the 1L PtSe$_2$ gap from ab-initio calculations.

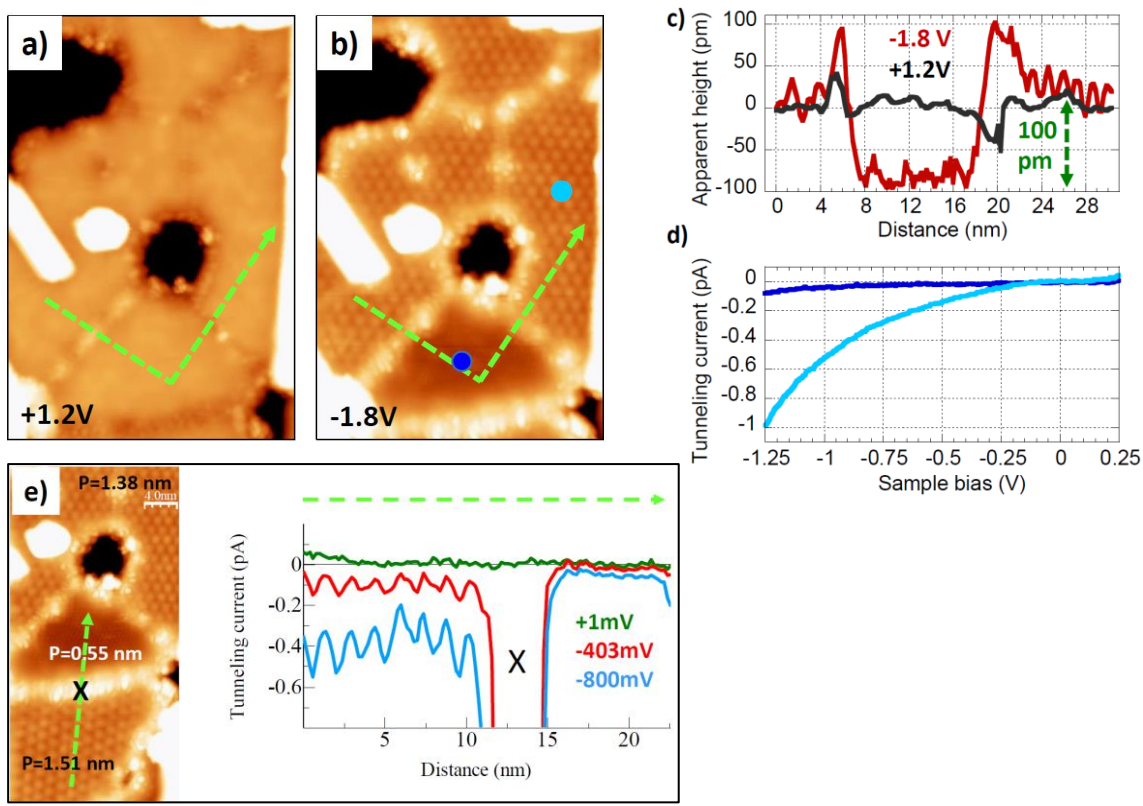

Figure S3: Moiré patterns and in-gap states at the 1L PtSe$_2$/BLG interface. a) and b): image of the same area with size 26x39 nm², taken at +1.2V (a) and -1.8V (b). The Z range (difference between black and white) is 0.53 nm for both images. c) Profiles taken along the path indicated by the dashed green arrow in a) (black curve) and b) (red curve). d) Bias dependence of the tunneling current measured within the 1L PtSe$_2$ bandgap on the small moiré (dark blue curve) and on the large moiré (light blue curve). The set-point bias/current for the spectra was +1.2V/500pA. e) Spatial variations of the tunneling current within the 1L PtSe$_2$ bandgap. Size of the image: 20x40 nm², sample bias: -1.8 V. A series of spectra was taken along the green dashed arrow (using setpoint: +1.2V/500pA), and the variation of the tunneling current as a function of the tip position is plotted for three different biases: +1 mV (green), -403 mV (red) and -800mV (light blue). The states contributing to the current are located inside the 1L-PtSe2 bandgap. The line for +1 mV gives the effective reference for "zero current". The label "**X**" indicates the boundary between the large and the small Moiré domains.

Figures S3-a) to S3-d) display data which are similar to the ones shown in Figure 1 (main text), but taken at another spot where large and small Moiré patterns (MPs) are visible. The periods P of the superstructures are indicated on Figure S3-e. One notices that the MPs remain visible in occupied state images taken at sample bias -1.80V (just below the VBM), and that the domain with the small period superstructure (P=0.55 nm) has lower apparent height (therefore, it appears as darker) than the ones where the period is larger (P>1 nm). Once again the MPs are absent at sample bias +1.2V, where all islands have the same apparent height. The curves shown in Figure S3-e) show that the tunneling current measured within the 1L-PtSe$_2$ bandgap is strongly modulated with the period of the Moiré on the large MP (located at Distance <10 nm), as expected (the setpoint bias for the spectra was +1.2 V, where no MP corrugation is seen). On the small MP (for Distance >15 nm) the modulations are much weaker, and the current is smaller than the minimum value recorded on the large MP (in absolute value and for a given bias). The data of Figure S3 lead to the same conclusions as those of Figure 1 in the main text.

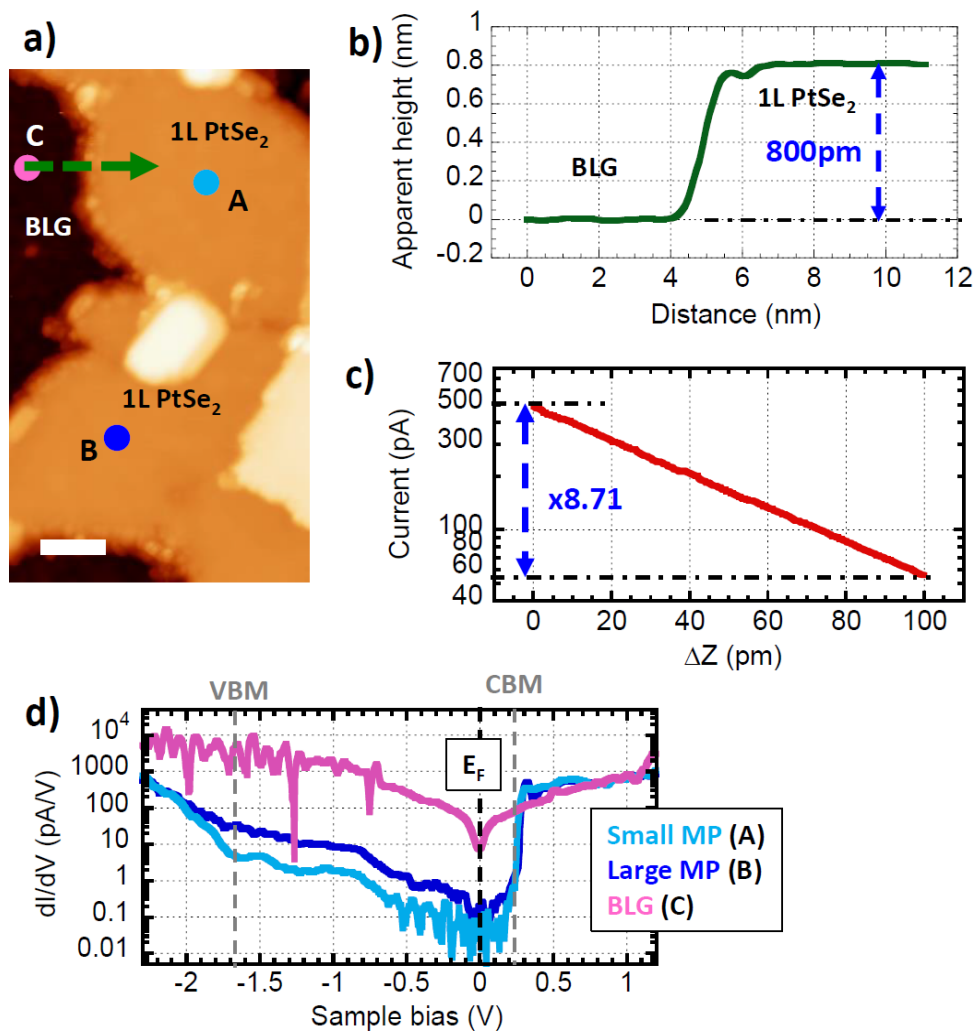

Figure S4: Additional data for Figure 1 of the main text. a) Same image as in Figure 1-a without contrast enhancement on the 1L PtSe$_2$ layer. Sample bias/current: +1.20 V/20 pA. Size: 23x40 nm², scale bar: 5 nm. b) Topographic profile along the green dashed line in a). The apparent height of the 1L PtSe$_2$ islands (relative to the BLG substrate) is 800pm for the sample bias +1.20V. c) Variation of the tunneling current as a function of the (relative) tip to sample distance ΔZ, measured at the sample bias +1.20V. The current vs. distance (I(Δz)) curve is perfectly exponential, and the current decreases by almost one order of magnitude (more precisely by a factor 8.71) for an increase by 100 pm of the tip-sample distance. This is a standard result for an I(Δz)

measurement in STS. d) Conductance spectra measured on the different domains shown in a): BLG substrate (point C), small period MP (point A) and large period MP (point B). The curves are displayed in log scale and extend over the PtSe$_2$ valence and conduction bands. Within the TMD bandgap the conductance (dI/dV signal) is at most 3 orders of magnitude larger on BLG than on the PtSe$_2$ islands. The dip at zero bias on the BLG curve is an ubiquitous feature of the spectra for graphene on SiC, that is **not** related to the Dirac point [23].

Figure S4 presents the measurements from which we have extracted the numerical values used when discussing the current vs. voltage (I(V)) curves shown in Fig. 1-d. For those measurements (Fig. S4-b), the tip to BLG (substrate) distance increases by 800 pm when the tip moves from a bare BLG area to a 1L PtSe$_2$ island (the spectra of Fig. 1-d were indeed taken with a setpoint bias +1.20V, for which all 1L PtSe$_2$ islands have the same apparent height). This is equivalent to ΔZ=800 pm for the tip-BLG substrate distance. According to Figure S4-c, this should result in a decrease by a factor $(8.71)^8 \approx 3.3 \cdot 10^7$ of the direct tunneling from the tip into the BLG. Although a rough estimate, this factor is orders of magnitude larger than the ratio of the tunneling currents or tunneling conductances (measured within the TMD bandgap) between bare BLG and 1L PtSe$_2$ islands, which amounts at most to $10^3$ from Fig. 1-d and Fig. S4-d. This indicates an "amplification" of the BLG states (which exist in the energy range of the 1L PtSe$_2$ bandgap) through the TMD layer, or equivalently an increase of the LDOS within the 1L PtSe$_2$ of the graphene derived in-gap states (compared to case of bare BLG at the same distance from the substrate). Hence, the graphene-derived electronic states acquire a partial PtSe$_2$ character within the TMD bandgap. In other words, there is a small admixture of states originating from PtSe$_2$ into the mainly graphene-based in-gap states. This is a strong indication for the existence of hybridized Graphene-PtSe$_2$ electron states (with mostly a BLG character) within the 1L PtSe$_2$ bandgap. The variations of the current (or of the conductance) detected within the TMD bandgap between the two domains with small and large MPs suggest that this hybridization depends on the period of the MP, and eventually on the twist angle between 1L PtSe$_2$ and BLG. Moreover, since all 1L PtSe$_2$ islands have the same apparent height at +1.2V sample bias (see Fig. 1, S1, S3, S4), the variations of the current within the TMD bandgap in the spectra of Figs. 1-d and S3-d (taken with set point voltage of +1.20V) lead to different apparent height of the MPs. This is clearly seen in Fig. 1: the much larger current measured at -0.80V on the large MP in Fig. 1-d corresponds to a larger apparent height of the island in Fig. 1-b (and Fig. 1-c) in the image taken with sample bias -0.80V (compared to the island with the small MP). Differences in the apparent heights of islands in images taken with sample biases located within the TMD bandgap thus also indicate a different degree of hybridization with the BLG states.

Such interfacial hybridization of graphene based in-gap states is not specific to the 1L PtSe$_2$/BLG system considered here. For instance, in the case of the interface between semiconducting Cr$_2$Ge$_2$Te$_6$ (CGT) and graphene, the admixture of CGT states in the in-gap states of the coupled system has been evaluated to about 2% close to the Dirac point [24].

Notice that this "amplification" of the tunneling current between the tip and the BLG substrate by the 1L PtSe$_2$ layer (or the occurrence of an hybridization between BLG and 1L PtSe$_2$ states at the interface) also leads to the finite height of the 1L PtSe$_2$ islands measured in constant current images acquired à -0.80V sample bias. For instance, the apparent height of the island shown in Fig. S4-a (and S4-b) decreases to 500 pm for a sample bias of -0.80V (not shown). This is consistent with the reduction of the current discussed in the previous paragraph and in connection with Fig. 1 in the main text.

**SI4 : Origin of the triangular superstructures (Moiré patterns) at the 1L-PtSe$_2$/BLG interface.**

In this section, we present data, which show that the triangular superstructures (with period P) observed inside the 1L PtSe$_2$ domains are related to the rotation angle θ between the crystallographic directions of the lattices of the TMD and of the BLG. We also show that the TMD band onsets at the 1L-PtSe$_2$/BLG interface do not depend on θ (i. e. on the value of P).

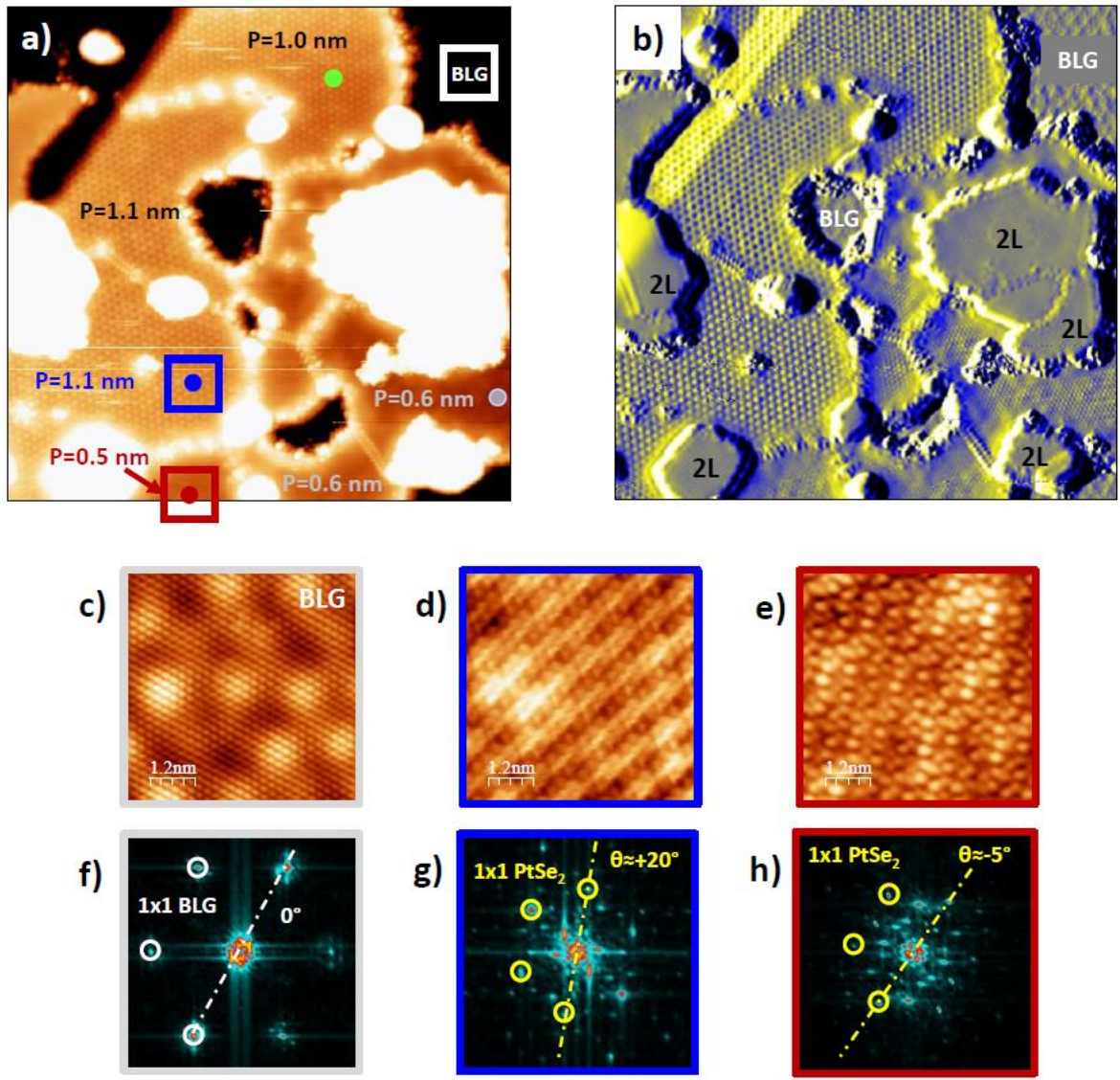

Figure S5: Rotational domains in 1L-PtSe$_2$. a) Constant current image taken with V$_s$=-1.80V and I$_t$=20 pA. Size: 60x60 nm². The periods P of some superstructures are indicated. The contrast has been enhanced on the 1L PtSe$_2$ phase. b) Same image as in a), but displayed in the derivative (dz/dx) mode, without any contrast enhancement on the 1L PtSe$_2$ layer. This helps better visualizing the MPs with small periods which develop in the domains which appear as dark in a). c), d) and e) Zoomed-in constant current images taken at selected areas on a). Images size: 6x6 nm², sample bias: +0.10V for the three images. The location of the images are indicated by squares of the same color in a). c) is for BLG (white square in a)), d) (e)) is for a large (small) MP on a 1L PtSe$_2$ island at the location indicated by a blue (red) square in a). f), g) and h) Fourier Transforms (FTs) of STM images displayed in c), d) and e) respectively. The first order spots of the reciprocal lattice are indicated by white circles in f) for BLG and by yellow circles in g) and h) for 1L PtSe$_2$. The value of θ for the different 1L PtSe$_2$ domains are estimated from the rotation between the white dashed line in f) and the yellow dashed lines in g) and h).

We present here measurements revealing the existence of rotational domains at the interface between 1L-PtSe$_2$ and BLG, and we illustrate the method we have used to measure directly the rotation angle θ. Although many of the islands correspond to θ≈0° as in Fig. S2, we have commonly found islands with large values of θ. This is explicitly shown in Fig. S5. Figure S5-a displays an area where several interconnected 1L PtSe$_2$ islands exhibit superstructures with different periods P. Zoomed-in images with atomic resolution have been taken with sample bias +100mV on these islands, as well as on the BLG substrate. Some of them are presented in Fig. S5-c to S5-e, and their Fourier transforms (FTs) are displayed in Fig. S5-f to S5-h. The spots corresponding to the atomic lattice are indicated by white (on BLG, Fig. S5-f) and yellow (on PtSe$_2$, Fig. S5-g and S5-h) circles. From these data one can get a direct estimate of the value of θ (with an accuracy of ±2°) for the different domains by measuring the angle between the white dashed line in Fig. S5-f and the yellow dashed lines in Fig. S5-g (θ≈+20°) and S5-h (θ≈-5°). From such measurements, we conclude that small values of P (P< 1 nm) correspond to small values of θ (θ<10°). Conversely, larger values of P correspond to rotation angles θ≥20°.

Beside those of the PtSe$_2$ lattice, the FTs diagrams in Fig. S5-g and S5-h reveal other spots that correspond to the triangular superstructures observed in the domains. As discussed in the main text and in section SI5 below, these superstructures are Moiré Patterns (MPs) which depends on the value of θ.

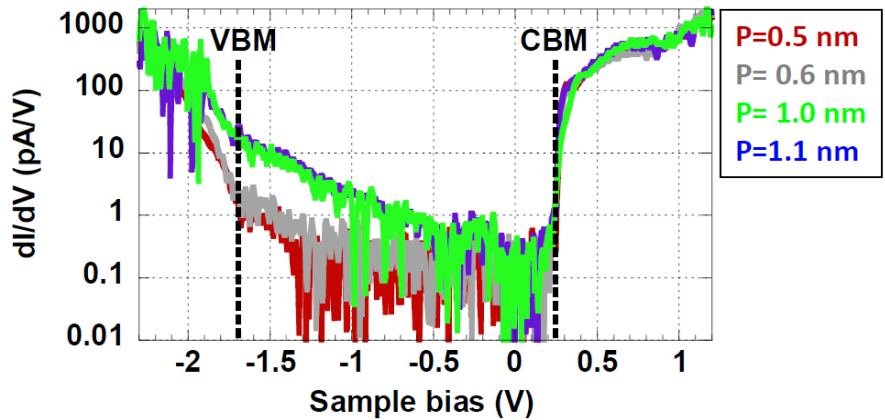

Figure S6: Band onsets at the 1L-PtSe$_2$/BLG interface. a) Tunneling spectra measured on different domains in Fig. S6-a using setpoint $V_s$=+1.20 V and $I_t$=500 pA. The locations of the spectra are indicated in Fig. S6-a by dots with the same color as the curves in Fig. S7. The values of the periods P of the related superstructures are given in the box on the right. The interface signal within the TMD bandgap increases strongly for large MPs (P≥1 nm). Otherwise, the positions of the CBM and of the VBM seem to depend only weakly on P, and thus on θ.

The tunneling spectra of Figure S6 indicate that the conductance measured within the TMD gap (in the region located between the lines labelled "VBM" and "CBM" in Fig. S6) is indeed larger in domains with large values of P (thus for large values of θ), which confirms the data of Fig. 1-d, S3-d and S4-d. Moreover, these spectra show that the positions of the valence band maximum (VBM) and conduction band minimum (CBM) of 1L-PtSe$_2$ at the interface with BLG are almost independent of θ (of P). From a technical point of view this later point is relevant since, in ab initio calculations, the band onsets for various SC TMDs on graphene may exhibit significant angular variations [24, 30]. This effect is presumably related to the strain induced by the choice of the commensurate cell [30]. These energy shifts may in turn (artificially) affect the hybridization of the graphene states close to the Dirac point with the TMD bands in the calculations [30].

As quoted before, it appears that the position of the CBM/VBM is essentially independent of the value of P (or θ). This indicates that the position of the Dirac point (DP) of the BLG substrate is also independent on θ. Indeed, a shift of the DP of the BLG (relative to the Fermi level) should induce a shift (with comparable amplitude) of the CBM of the 1L PtSe$_2$ layer owing to the absence of Fermi level pinning at the interface (see [21, 22] and references therein). No significant energy variations of the band onsets as a function of P (or θ) are observed in the data (see Figs. S4-d and S6). Therefore, we expect the Dirac point of BLG to remain essentially at the same energy for all rotation angles.

**SI5 : Interpretation of the Moiré patterns at the 1L-PtSe$_2$/BLG interface (additional data for Figure 2).**

In this section, we show:

- In Figure S7 images taken at some of the same spots as in Fig. 2, but at different biases, to show that the MPs are observed at various biases within the gap. We also provide an image for the reference BLG substrate.
- In Figure S9 images taken at different biases (+0.10V and -0.80V) on different MPs corresponding to rotation angles θ between 4° and 20°, to show that one type of MP, namely MP2, exists at any bias over the whole range of values of θ.
- In Figure S10 the calculated values of the period Pi of the MPi and of the angle αi between the axes of MPi and of the 1L PtSe$_2$ layer for MP1 (i=1) and MP2 (i=2).

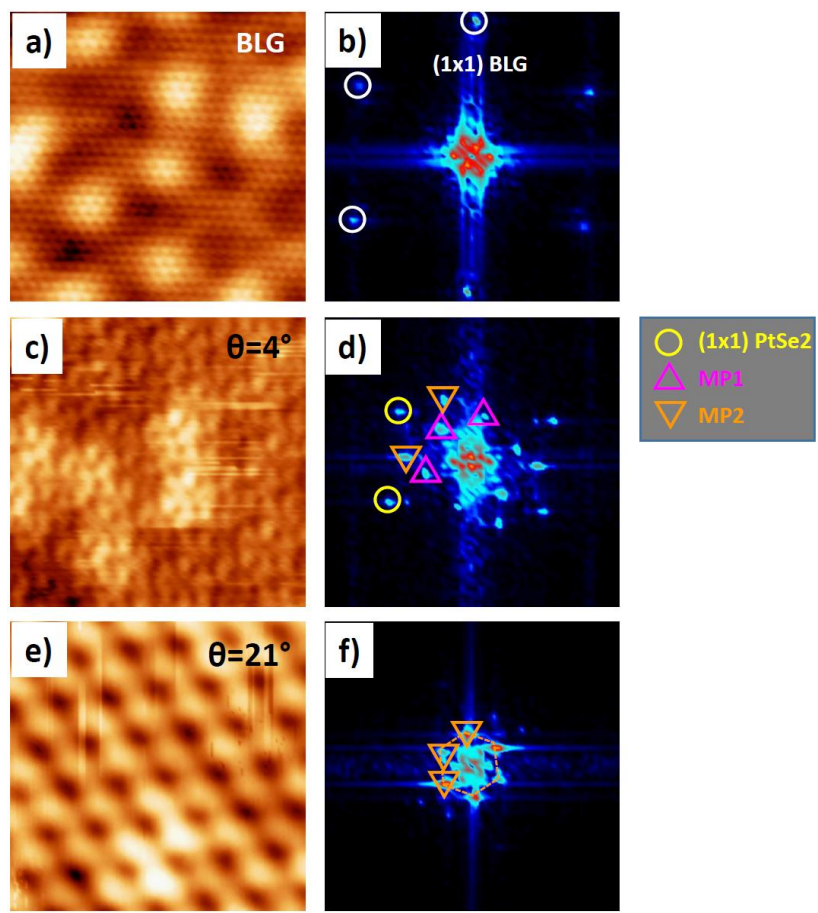

Figure S7: Moiré patterns. a) Constant current image of the BLG surface and its Fourier Transform (FT) b) to set the reference of substrate orientation for the data in Fig. 2 and Fig. S7. Sample bias: +0.50V. c) Constant current image and its FT d), taken at the same location as in Fig. 2-c, but with sample bias -1.20V. The two MPs remain visible for this bias corresponding to in-gap states. e) Constant current image and its FT f) taken at the same location as in Fig. 2-e, but with sample bias -0.80V. The MP2 features remains clearly visible, although the atomic resolutions is lost (and thus possible replicas have disappeared in the FT). MP1 is not visible in e) and f). Size of the images: 6x6 nm². Notice that the images are rotated by 90° compared to Fig. S5 and Fig. S8.

Figures S7 and S8 are intended to show that the MPs (MP2 and MP1 when present) remain observable with the same period for all biases corresponding to states located within the TMD bandgap, or close to the VBM (-1.8V).

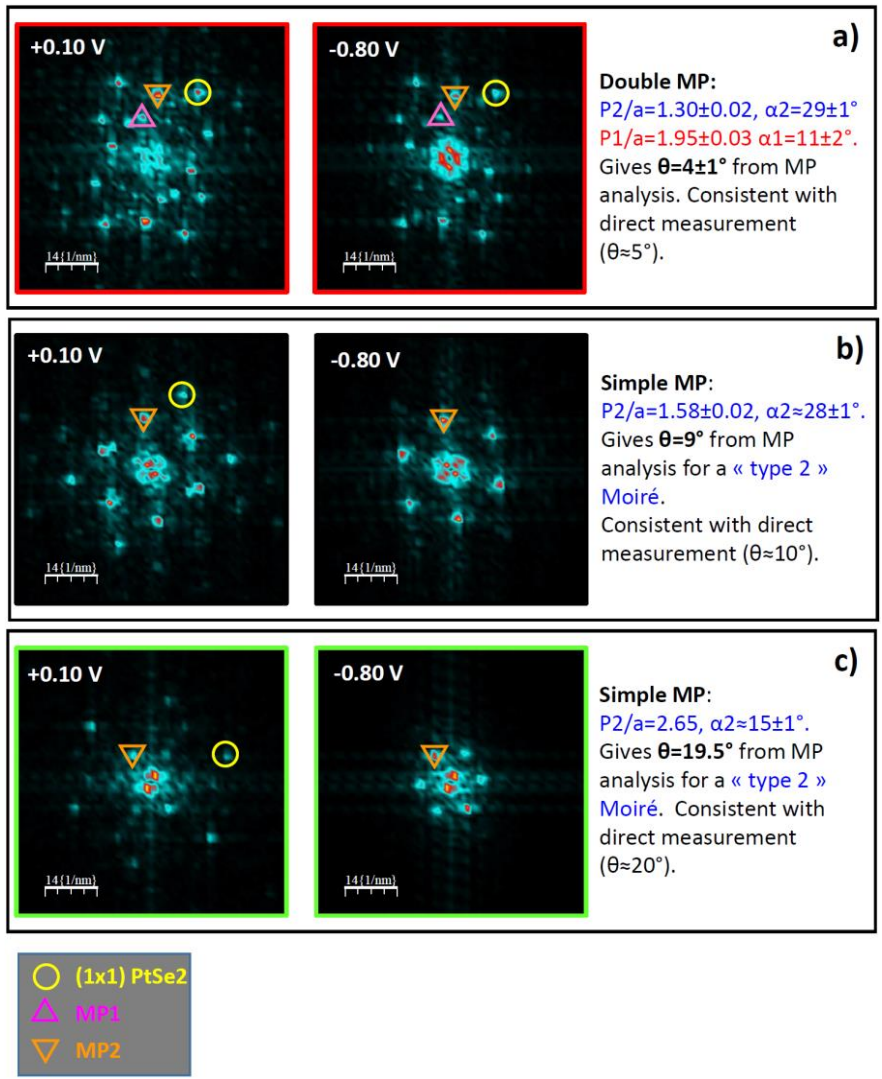

Figure S8: Fourier Transform (FTs) of images with size 4.5x4.5nm² taken at sample biases +0.10V and -0.80V on different locations of Figure S6. The reference axes for the BLG substrate are as in Fig. S6. The period Pi of the MPi (in units of the PtSe$_2$ lattice constant a) and the angle αi (see Fig. S9-d) between the main directions of MPi and of the 1L PtSe2 layer for MP1 (i=1) and MP2 (i=2) as measured on these FTs are indicated. The values of θ resulting from the analysis of the MPs are in good agreement with the direct measurements such as the ones shown in Fig. S5. Notice that in panel a) the value θ=4±1° is obtained from the analysis of either MP1 or MP2 (consistent with both of them).

The underline{important point here} is that MP2 is clearly observed for any value of the rotation angle θ between PtSe$_2$ and BLG, which is not the case for MP1. Some weak structures observed around the spots of the (1x1) PtSe$_2$ lattice at +0.10 V in Fig. 2-f and Fig. S8-c can be regarded as replicas from MP2 around the (1x1) spot of the atomic lattice and/or as spots of MP1 (see next paragraph). These structures disappear at large negative bias (-0.80 V) in Fig. S7-f and Fig. S8-c, and only the MP2 spots remain visible.

Technically, the reciprocal lattice vectors MP1* and MP2* of the Moiré Patterns MP1 and MP2 are related by the reciprocal vector of the PtSe$_2$ lattice b*$_{TMD}$. This is shown in Fig. 2-g and S9-c below. It means that MP1 spots can appear as replica of the MP2 spot around the (1x1) spots of the PtSe$_2$ lattice (and vice-versa). The appearance of such "replicas", which show up in the FTs of STM images as additional spots shifted by reciprocal lattice vectors from the primary ones is a mere consequence of the Bloch character of the wavefunctions of the electronic states which contribute to the tunneling current [25].

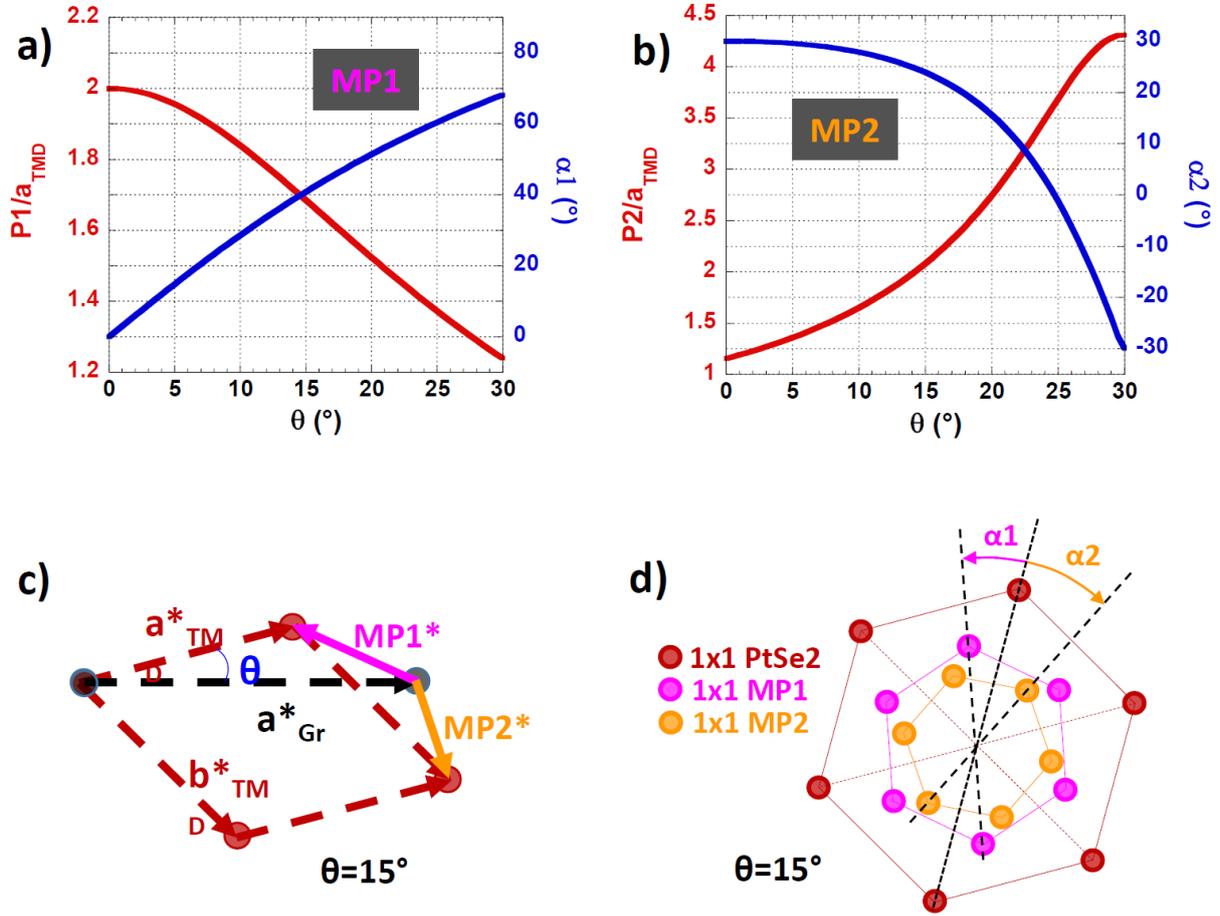

Figure S9: Calculation of the pseudo-period Pi of MPi (in units of the TMD lattice constant $a_{TMD}$) and of the angle αi (modulo 60°) between the main directions (axis) of the MPi and of the PtSe$_2$ lattice for the two MPs MPi (i=1, 2) as a function of the rotation angle θ between 1L PtSe$_2$ and BLG. a) for MP1, b) for MP2. The FTs of Fig. S9 and of Fig. 2 (main text) were analyzed using these plots. c) Construction used to derive the reciprocal vectors MP1* and MP2* of the moirés patterns MP1 and MP2. d) Relative location in the reciprocal space of the first order spots of the PtSe$_2$ lattice (in red), of the MP1 moiré pattern (in pink) and of the MP2 moiré pattern (in orange). This diagram in constructed from the drawing in c), both are for θ=15°. In the analysis of the experimental data, the values α1 and α2 were measured from the FTs of the images as indicated in d).

The model used to calculate the data for the curves displayed in Fig. S9-a and S9-b is the one from [26] that we have already used in Ref. [27] to analyze the MPs at a graphene-substrate interface. The FTs of Fig. S8 and of Fig. 2 (main text) were analyzed using these plots. For the calculations, we have assumed that the lattice constants of 1L PtSe$_2$ and of BLG were commensurate for θ=0°, with 3x$a_{Gr}$=2x$a_{TMD}$, which is a good approximation (3x$a_{Gr}$=3x0.246nm≈0.74nm; 2x$a_{TMD}$=2x0.375nm=0.75nm, $a_{Gr}$ and $a_{TMD}$ being the in-plane lattice constants of BLG and 1T-PtSe$_2$ respectively).

**SI6 : Moiré pattern and interlayer coupling (additional data for Figure 3).**

In this section, we show:

- In Figure S10 the construction that allows identifying the vector which connects the points k1 and k2 in Figure 3-a to 3-c to the reciprocal lattice vector MP2* defined in Fig. 2-g and in Fig. S9-c.
- In Figure S11 the results of ab-initio calculations which indicate that the hybridization scheme at the interface between monolayer graphene (SLG) and 1L PtSe$_2$ should be very similar to the one analyzed here for the interface between BLG and 1L PtSe$_2$.
- In Figure S12 a comparison of the band structure obtained from ab-initio calculations for the interface between 1L PtSe$_2$ and monolayer graphene for θ=0° and for θ=19.1°. This figure presents another way to estimate the evolution of the coupling strength for large rotation angles.
- We add technical remarks for the reading of Fig. 3-d in the main text.

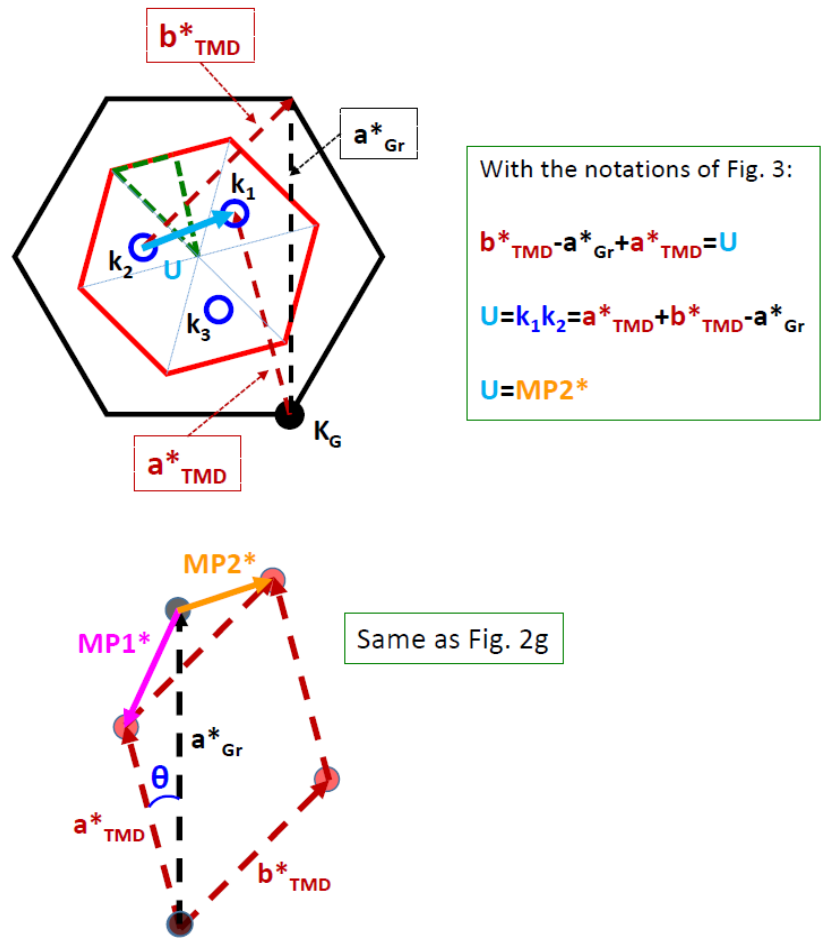

Figure S10: Relation between the hybridization at the interface and the Moiré pattern MP2. From the coupling scheme considered in Ref. [21, 22], the states close to the K point of graphene (labelled K$_G$) are coupled to the states **k$_1$**, **k$_2$** and **k$_3$** of 1L PtSe$_2$. These **points (ki's)** correspond to the backfolding (via reciprocal lattice vectors of 1L PtSe$_2$, labelled **a*$_{TMD}$** and **b*$_{TMD}$** here) in the first Brillouin zone of 1L PtSe$_2$ of the three equivalent K$_G$ points of graphene (which are connected by graphene reciprocal lattice vectors such as **a*$_{Gr}$**). From the construction shown above, it is obvious that the vector **U** connecting **k$_1$** to **k$_2$** is the reciprocal lattice vector **MP2*** of the Moiré pattern MP2. This also holds for the other couples of points (**k$_1$**,**k$_3$**) and (**k$_2$**,**k$_3$**), as well as for the 3 points (**k'$_1$**, **k'$_2$**, **k'$_3$**) coupled to the states close to the K$_G$' (or –K$_G$) point of graphene.

The coupling scheme illustrated in Fig. S10, following Ref. [28, 29], is valid not only at the graphene K point but also in its neighborhood. Thus, the three states of 1L PtSe$_2$ hybridized to any graphene state in the vicinity of the K point (but not exactly at the K point) remain coupled by the reciprocal lattice vectors of MP2.

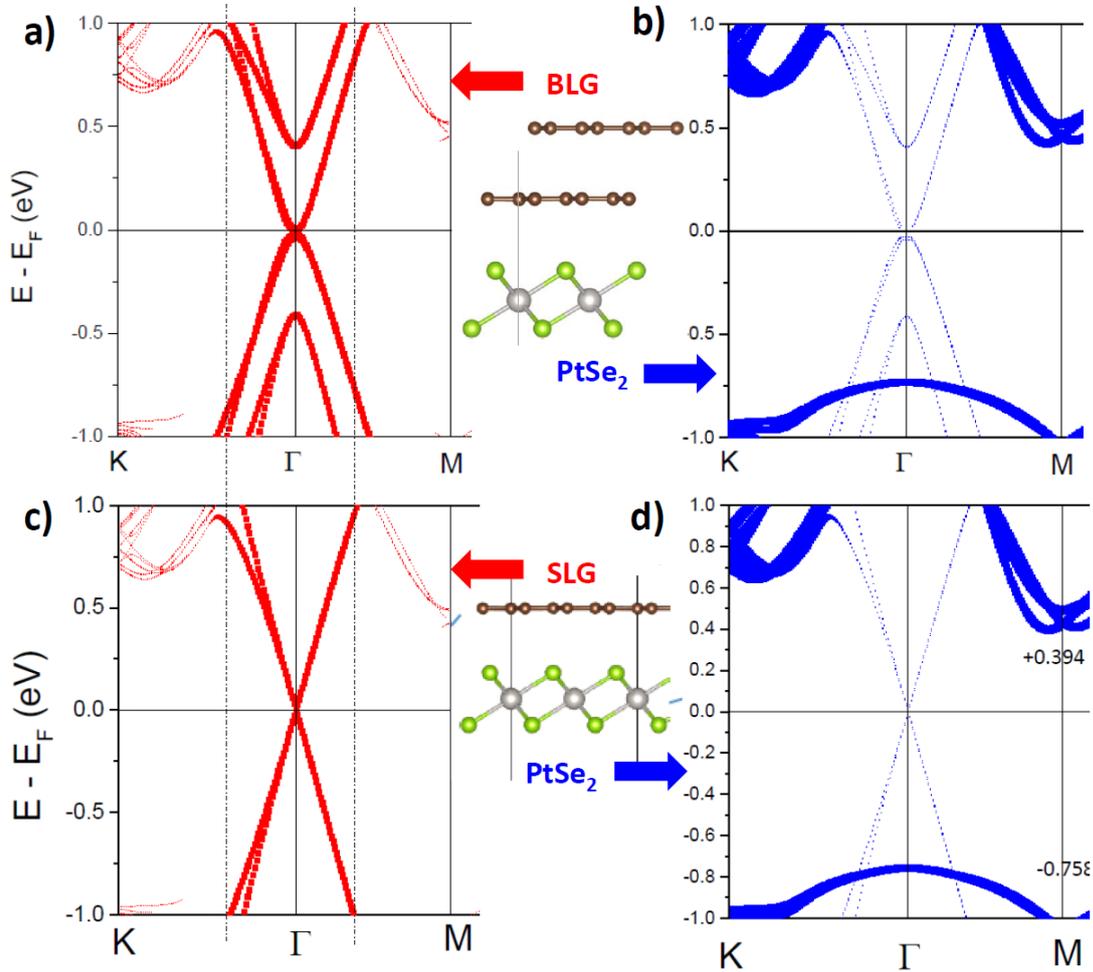

Figure S11: Calculated electronic structure for the interface between 1L PtSe$_2$ and BLG (in a) and b)) and between 1L PtSe2 and monolayer graphene (SLG, in c) and d)). The relaxed structure of the system is shown in the central panel. The common (commensurate) unit cell is (2x2) for PtSe$_2$ and (3x3) for BLG or SLG. The atomic lattices of SLG or BLG and 1L PtSe$_2$ are aligned (i. e. θ=0°). a) and b) display the projection of the band structure of the heterostructure on the BLG (red) and on the 1L PtSe$_2$ (blue) respectively (same as in Fig. S5). c) and d) display the projection of the band structure of the heterostructure on the SLG (red) and on the 1L PtSe$_2$ (blue) respectively. In a) to d), the size of the symbols (dots) is proportional to the weight of the site-projected state in the corresponding layer.

We refer to Fig. 1-e and Fig. 1-f for the interpretation of the Fig. S11. From the projected band structure on the 1L PtSe$_2$ (blue curves), it is clear that states that reproduce the low energy band structure of BLG and SLG have some (small but finite) weight within the TMD bandgap. We thus expect that the states close to the Dirac point in monolayer graphene hybridize with states in 1L PtSe$_2$ in the same way as for the BLG case considered in the present work.

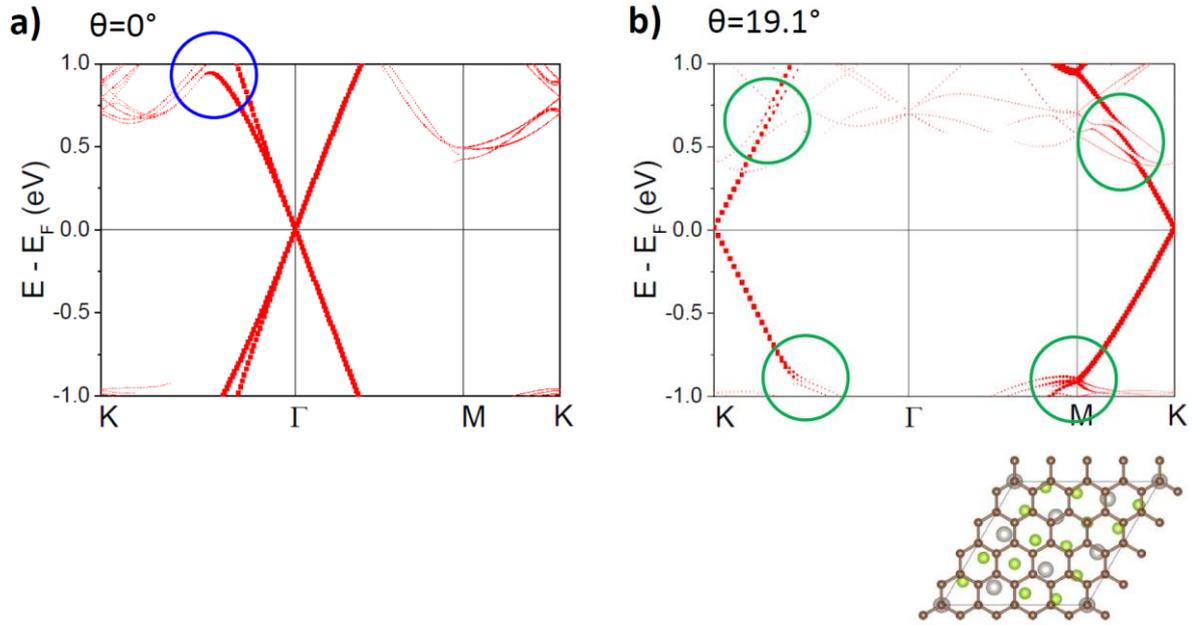

Figure S12: Band structure projected on the graphene layer for 1L PtSe$_2$/SLG twisted bilayers for two values of the rotation angle θ. a) θ=0°, same Figure as S12-c, corresponding to an aligned bilayer. b) θ=19.1°, corresponding to a large rotation angle. The top view of the commensurate structure for θ=19.1° is shown in the lower part of the panel. In a) and b), circles highlight "anticrossings" between the graphene and TMD bands, indicating a significant coupling of the states.

In the main text (Fig.3-d) we interpret the angular variations of the hybridization between the TMD and graphene states in terms of the energy difference between the levels in the uncoupled layers only, following e.g. Ref. [28]. This qualitative interpretation has been questioned based on more recent ab initio calculations [24, 30]. The authors suggest instead to use the "anticrossings" that show up in the projected band structure of the coupled system to evaluate the strength (the efficiency) of the hybridization between the states of the two layers. We present in figure S13 the results of our ab initio calculations for the case of 1L PtSe$_2$ on monolayer graphene.

Up to now (Fig. 1-e and 1-f, Fig. S11) we have shown the band structure for aligned bilayers, thus for θ=0°. In this case a commensurate unit cell consists of three unit cells of graphene and two unit cells of PtSe$_2$, with a small lattice mismatch (<2%). This corresponds to P/$a_{PtSe2}$=2.0 for the MP1 (Fig. S9). We have found another quasi-commensurate unit cell for a large rotation angle θ=19.1° and with a similarly small lattice mismatch (<1%). As shown (among others) in Ref. [30], the lattice mismatches and thus the strain effects have an impact on the results of ab initio simulations. The supercell used in the calculation is displayed at the bottom of Figure S12-b. Its size is four times the unit cell of graphene. This corresponds to P/$a_{PtSe2}$=2.67 as expected for a MP2, see Fig. S9.

The anticrossings between the graphene and the 1L PtSe$_2$ bands located within ±1.0 eV from the Dirac point (DP) are highlighted in Fig. S12 by blue circle for θ=0° and green ovals for θ=19.1°. For θ=0° there is only one anticrossing, located about 1.0 eV above the DP (Fig. S12-a). For θ=19.1° (Fig. S12-b), two anticrossings occur at lower positive energy (between +0.5 eV and +1.0 eV above the DP). We also notice two additional anticrossings in the valence band, around -0.8 eV or -0.9 eV below the DP. Since the coupling between the graphene and TMD bands occur at lower energy from the DP in the case of the large rotation angle, we expect a stronger hybridization for the in-gap (interfacial) states in this later case. This consideration thus leads to the same conclusion as the one based on the energy difference between the uncoupled states considered in the main text. By the way, from Fig. S12, the argument using the energy difference between SLG and TMD states also holds for θ=19.1° compared to the case of aligned layers (θ=0°). In our case, these qualitative arguments are both consistent with the experimental findings.

**Technical remarks for the reading of Fig. 3-d of the main text.**

- The values of the $k_i$'s (i=1, 2 or 3) given in this figure for different values of θ are only valid for bilayer graphene states located exactly at the $K_G$ point. The $k_i$'s are equivalent only in this case due to threefold symmetry. For a BLG state slightly off the K point, the energy distance to the 1L $PtSe_2$ bands should be slightly different for three $k_i$'s. This is anyway a second order effect.

- Using the conventional representation of the band structure in Fig. 3-d, only the values of θ for which the $k_i$'s are located on the main directions ΓM, ΓK or KM of the TMD BZ can be taken into consideration. For θ=30° the $k_i$'s are on the ΓM line, actually close to Γ (at 0.27xΓM). For θ=0° they are on the KM line, but owing to the (quasi) commensurability between a (2x2) of $PtSe_2$ and a (3x3) of BLG they are indeed located at (or very close to) the M point. The angle (θ=24.7°) for which the ki's fall on the ΓK direction (and its position at 0.28xΓK on this line) has to be computed for each TMD [28].